\documentclass{ws-ijmpb}

\begin{document}

\markboth{\'Alvaro Corral}
{Statistical Tests for Scaling 
in
the Inter-Event Times of
Earthquakes in California 
}

%
\catchline{}{}{}{}{}
%

\title{STATISTICAL TESTS FOR SCALING 
IN
THE INTER-EVENT TIMES OF
EARTHQUAKES IN CALIFORNIA 
}

\author{\'ALVARO CORRAL}

\address{Centre de Recerca Matem\`atica,
Edifici Cc, Campus UAB,
E-08193 Bellaterra, Barcelona, Spain
\\
ACorral at crm dot es
}

\maketitle

\begin{history}
\received{Day Month Year}
\revised{Day Month Year}
\end{history}

\begin{abstract}
We explore in depth the validity of a recently proposed 
scaling law for earthquake inter-event time distributions
in the case of the Southern California, using
the waveform cross-correlation catalog of Shearer {\it et al.} 
Two statistical tests are used: on the one hand, the standard
two-sample Kolmogorov-Smirnov test is in agreement 
with the scaling of the distributions.
On the other hand, the one-sample
Kolmogorov-Smirnov statistic complemented with Monte Carlo 
simulation of the inter-event times, as done by Clauset {\it et al.},
supports the validity of the gamma distribution as a simple
model of the scaling function appearing on the scaling law, 
for rescaled inter-event times above 0.01, except for the 
largest data set (magnitude greater than 2).
A discussion of these results is provided. 
\end{abstract}

\keywords{Statistical seismology; scaling; goodness-of-fit tests; complex systems.}

\section{Introduction}

In the last years considerable attention has been 
addressed to the distribution of inter-event times 
in natural hazards, in particular earthquakes%
\cite{Bak.2002,Corral_pre.2003,Corral_physA.2004,Corral_prl.2004,Bunde,Astrom,Davidsen_fracture,Geist_Parsons,Baiesi_flares,Corral_fires}, 
but also in human responses\cite{Struzik,Osorio,Corral_words} 
and social behavior\cite{Barabasi,Yamasaki,Harder_Paczuski,Goh}.
In some of these systems, the shape of the inter-event-time distribution
for events above a certain threshold in size
is independent on the threshold.
Indeed, let $\tau$ denote the inter-event time, 
defined as the time between consecutive events
above a size threshold $s$ and
let $D_s(\tau)$ be its probability density, 
then we can write
\begin{equation}
D_s(\tau) = R_s f(R_s \tau),
\label{scaling}
\end{equation}
where $f$ is a scaling function that
provides the shape of $D_s(\tau)$
and
$R_s$ is the occurrence rate 
for events above $s$, providing the scale of $D_s(\tau)$
(and $R_s^{-1}$ provides the scale for $\tau$).

If the size distribution follows a power law
(which is not always the case\cite{Corral_fires}), 
then $R_s \propto 1/s^\beta$ 
(where $\beta$ is the exponent of the cumulative size distribution),
and then
\begin{equation}
D_s(\tau) =  \hat f(\tau / s^{\beta} )/s^\beta,
\end{equation}
which turns out to be a scaling law,
equivalent to those obtained in 
the study of critical phenomena\cite{Stanley_rmp}.
The law reflects a scale-invariant condition:
there exist a change of scale in $\tau$ and $s$ 
(a linear transformation) that does not lead to 
any change in the statistical properties of 
the process, at least regarding the inter-event-time probability
density.
The function $\hat f$ is just the scaling function $f$,
except for proportionality constants.

Why is this scaling law of some relevance or interest?
In general, when events are removed from a point process
(as it is done in our case by raising the size threshold),
the resulting inter-event-time distribution changes
with respect to the original one,
and a scaling law as Eq. (\ref{scaling}) does not apply.
However, 
for high enough thresholds $s$
(for the extreme events that are of interest in 
hazard assessment studies),
and when the events are randomly removed,
it is expected that the resulting time process
tends to a Poisson process
(which means that the occurrence of the extreme events 
is independent on the history of the process, 
and just some multi-faced dice, thrown in continuous time, 
decides if an event takes place or not).
From the point of view of statistical physics, 
the Poisson process constitutes a trivial 
fixed-point solution of the renormalization equations
describing the {\it thinning} or decimation
performed in event occurrence when the size threshold is raised%
\cite{Corral_prl.2005,Corral_jstat}.
For event occurrence on a large spatial scale,
as for instance worldwide earthquakes, there is a second reason
to expect exponential inter-event-time distributions:
the pooled output of several time processes
(i.e., China seismicity, superimposed to Japan seismicity, etc...)
tends to a Poisson process 
if the processes are independent\cite{Cox_Lewis}.

It is therefore surprising not only that the scaling function 
$f$ is not exponential,
but also that a non-exponential scaling function exists.
In the case of earthquakes\cite{Corral_prl.2004} 
(and fractures\cite{Astrom,Davidsen_fracture})
$f$ is approximated by the so called gamma distribution,
with parameters $\gamma$ and $a$,
\begin{equation}
f(x) = \frac 1 {a\Gamma(\gamma,m/a)} \left(\frac a x \right)^{1-\gamma}
e^{-x/a},
\, \mbox{ for } x=R_s\tau \ge m,
\label{gamma}
\end{equation}
where $\Gamma(\gamma,m/a)$ is the complement of the incomplete
gamma function (not normalized), 
$\Gamma(\gamma,u)\equiv\int_u^\infty u^{\gamma -1}e^{-u} du$.
The cutoff value $m$ is not considered a free parameter
but fixed and
the scale parameter $a$ is not independent
but can be obtained from the value of $\gamma$ and $m$ taking into account that
$\langle x \rangle = \langle R_s \tau \rangle =\int_m^\infty x f(x)dx =1$ 
(using that $R_s$ is the inverse of the mean inter-event time).
For stationary seismicity, as well as from fracture and nanofracture
experiments, the shape parameter $\gamma$ turns out to be close to 0.7,
see Refs. \refcite{Corral_prl.2004,Astrom,Davidsen_fracture}.
The reason to disregard $x-$values below $m$
is due, on the one hand, to the 
incompleteness of seismic catalogs on the shortest
time scales and to the existence errors in the determination
of the inter-event times when these are small, 
and on the other hand to the breakdown 
of the stationarity condition in those short time scales
by small aftershock sequences.

The usual way to establish the validity of a
scaling law such as Eq. (\ref{scaling}) 
is by plotting the different rescaled quantities together 
(in our case inter-event-time distributions for different thresholds)
and judge visually if they collapse onto 
a single curve or not.
It would be nice if one could put some numbers
into the quality of the scaling and the fit of the scaling function
and test their limits of validity.
Let us note that
Kagan has argued that one of the reasons
because theoretical physics has failed not only to predict but to 
explain earthquake occurrence is 
due to the poor use of statistics by the researchers in the field\cite{Kagan_calcutta}.
Indeed, ``the quality of current earthquake data statistical analysis is low.
Since little or no study of random and systematic errors is performed,
most published statistical results are artifacts.''
We believe this criticism has applicability beyond
the case of statistical seismology.

In this paper we will first use the Kolmogorov-Smirnov two-sample test
in order to evaluate the fulfillment of the inter-event-time scaling law 
(\ref{scaling}) in Southern California seismicity.
Next, the goodness of the fit of the scaling function (\ref{gamma})
to the rescaled inter-event-time densities will be tested
by adapting the procedure introduced by Clauset {\it et al.}\cite{Clauset},
consisting in maximum likelihood estimation of parameters,
Kolmogorov-Smirnov one-sample statistic evaluation,
and Monte Carlo simulation of the inter-event times in order to 
compute the distribution of the statistic.

\section{Data}

The seismological data used will be the Southern California
waveform cross-correlation catalog of Shearer {\it et al.}\cite{Shearer2}
(for which, as far as the author knows, no study has published 
plain inter-event-time distributions $D_s(\tau)$,
nevertheless, see also Refs. \refcite{Corral_prl.2006,Davidsen_Grassberger}).
The catalog spans the years 1984-2002 (included),
containing 77034 earthquakes with magnitude $M\ge 2$.
Notice that we will use $M$ as a measure of size, 
although by the Gutenberg-Richter law it is not
power-law distributed but exponentially distributed\cite{Kanamori_rpp}.
In order to recover a power-law distribution one has
to deal with the seismic moment, or the energy, 
which are exponential functions of the magnitude.

We will concentrate in earthquake occurrence under 
stationary conditions. 
It is well know that earthquakes trigger more earthquakes
with a rate that changes in time following the Omori law\cite{Kanamori_rpp}.
In general, this breaks stationarity, as the rate of 
occurrence is not constant in time; however, at (relatively) large scales
the resulting superposition of time-varying rates yields
a constant rate, as it happens in worldwide seismic occurrence, 
and also for Southern California in certain time periods
in which the largest earthquakes do not occur,
see Fig. 1 of Ref. \refcite{Corral_tectono}.
Precisely for this reason, inter-event-times for stationary seismicity are more
reliable than for non-stationary periods, 
as the large earthquakes present in the latter case 
prevent the detection of the small ones\cite{Helmstetter_Kagan_forecast},
which has dramatic consequences in the computation of
the inter-event times.

The stationary time periods under consideration in this paper are
(refining those in Ref. \refcite{Corral_tectono}, 
following Ref. \refcite{Corral_TN}):
$1984-1986.5$, 
$1990.3-1992.1$, 
$1994.6-1995.6$,
$1996.1-1996.5$, 
$1997-1997.6$, 
$1997.75-1998.15$, 
$1998.25-1999.35$, 
$2000.55-2000.8$,
$2000.9-2001.25$, 
$2001.6-2002$, 
$2002.5-2003$,
where time is measured in years,
1 year = 365.25 days (every 4 years an integer value
in years corresponds to the true starting of the year) .

\section{Testing Scaling}

A simple way to quantify the validity of the scaling hypothesis
in probability distributions can be obtained 
from the two-sample Kolmogorov-Smirnov (KS) test,
which compares two empirical distributions.
The procedure begins with the calculation
of the maximum difference, in absolute value,
between the rescaled cumulative distributions of the two data sets,
i.e.,
\begin{equation}
     d_{kl} \equiv \mbox{max}_{\forall x} |P_k(x)-P_l(x)|;
\end{equation}
as we have more than two data sets, 
we label them with indices
$k$ and $l$.
The empirical cumulative distribution functions
$P_k(x)$ are calculated as the
fraction of observations in data set $k$
below value x,
and constitute an estimation of the
theoretical cumulative distribution function,
$F_k(x) \equiv \mbox{Prob}[\mbox{variable } < x ]
=\int_m^x D_k(x) dx$.

Obviously, the difference $d_{kl}$ is randomly distributed, 
and therefore we can refer to it as a statistic.
The key element of the KS test is that 
when the data sets $k$ and $l$ come indeed from the same
underlying distribution $F(x)\equiv F_k(x)=F_l(x)$, 
the distribution
of the KS statistic $d_{kl}$ turns out to be independent
on the form of $F(x)$ and can be easily computed.
Therefore, the resulting value of $d_{kl}$ can be considered
as small or large by comparison with its theoretical distribution.
Under the null hypothesis that both data sets come from 
the same distribution,
the probability that the KS statistic is larger than the 
obtained empirical value $d_{kl}$ gives the so called $p-$value, 
which constitutes the probability of making an error 
if the null hypothesis 
is rejected.
The formulas for the probability distribution of
$d_{kl}$ are simple enough and are given by Press {\it et al.}\cite{Press},
depending only on the number of data $N$ in each of the sets;
so, approximately, for large $N_e$,
\begin{equation}
p=
\mbox{Prob [ KS statistic $ > d $ ]} = 
Q([\sqrt{N_e}+0.12 + 0.11/\sqrt{N_e}] d),
\label{Q}
\end{equation}
with $Q$ a decreasing function taking values between 1 and 0
(see Ref. \refcite{Press})
and $N_e$ an effective number of data points
(the ``reduced'' number, or one half of the harmonic mean
of the number of data).
Nevertheless, in order to calculate the $p-$value 
it is simpler to use the numerical routines provided
in the same reference\cite{Press} (in particular, the routine called
{\tt probks}).

Notice that we have to compare the distributions of 
seismicity for $M \ge M_k$ and $M \ge M_l$ after rescaling, 
i.e., as a function of $R_k\tau$
and $R_l\tau$, respectively (otherwise, without rescaling, the distributions
cannot be the same).
In order to do that, for each data set, we first calculate 
the mean value of the inter-event time, $\langle \tau \rangle_{k}=R_k^{-1}$,
and then, we disregard inter-event time values such that
$R_k \tau < m$. 
The elimination of the smallest values 
increases the mean value of the remaining rescaled inter-event times,
so, we repeat the procedure:
we recalculate the mean inter-event time and rescale 
again the data by the new rate, disregarding those values
below $m$. The resulting data set has a mean value very close to one.
We will assume that this procedure does not invalidate the applicability
of the formulas we use for the calculation of the $p-$value.

\begin{figure}[bt]
\centerline{\psfig{file=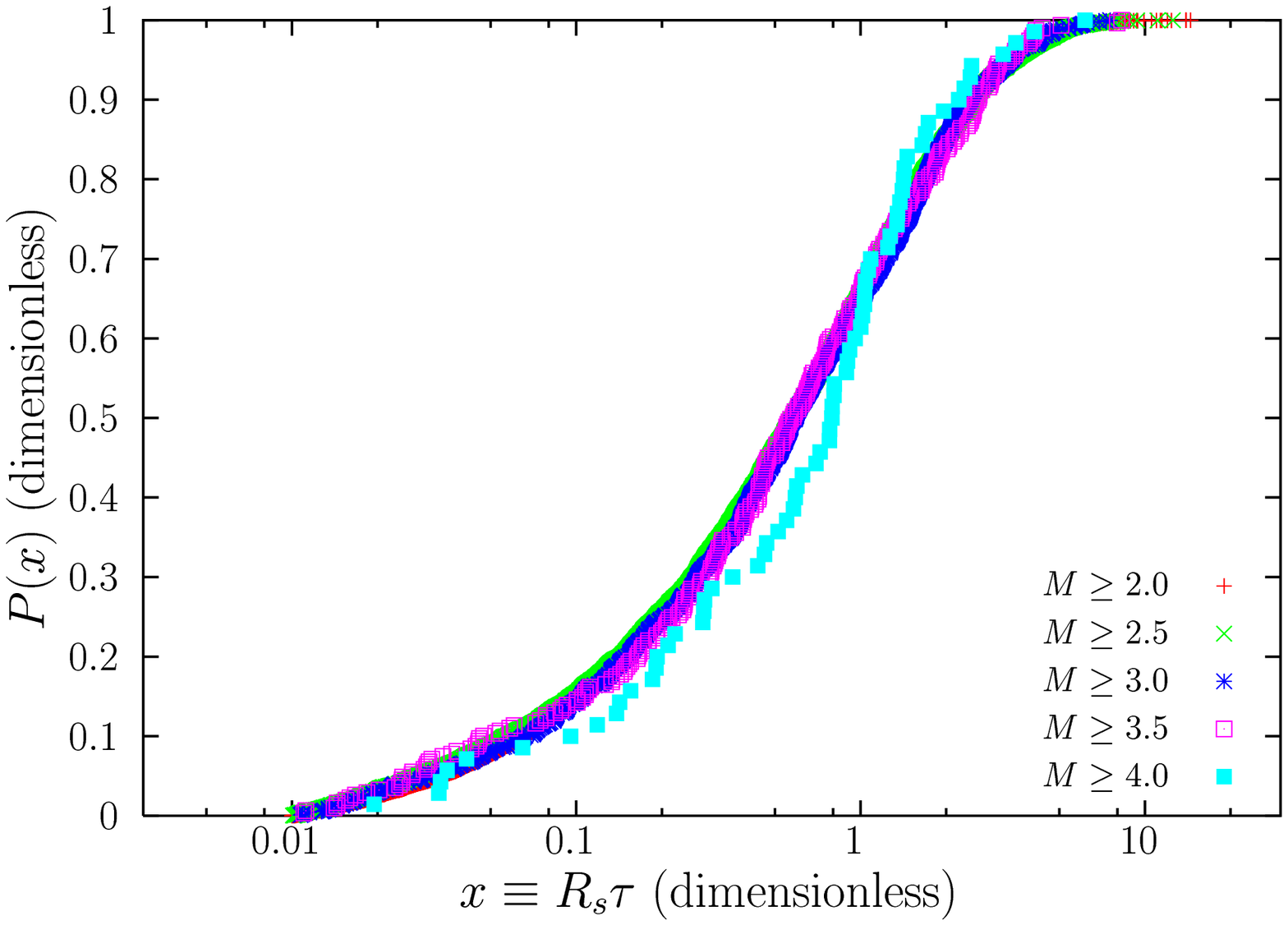,height=6cm}}
\vspace*{8pt}
\caption{Rescaled inter-event-time cumulative distributions
for Southern California stationary seismicity,
fixing minimum $x-$value $m=0.01$.
The collapse of the distributions 
is an indication of scaling, in agreement with the
results of the KS tests performed.
\label{twosample}
}
\end{figure}

The rescaled inter-event-time cumulative distributions for 
different magnitude ranges are shown at Fig. \ref{twosample},
ranging from $M\ge 2$ to $M \ge 4$, fixing $m=0.01$.
The scaling seems rather good, except for 
the case $M\ge 4$.
Table \ref{t1} shows the KS statistic for each pair of distributions,
as well as the corresponding $p-$values.
Due to their high values (in all cases larger than 0.18
but in some others larger than 0.95), 
the null hypothesis cannot be rejected 
and each pair of data sets are compatible
with the same underlying distribution,
and therefore we have to agree with the
scaling hypothesis (within statistical significance).
Let us note that the $p-$value, being itself 
originated by a random set, is a random quantity
(when different data sets are considered),
and it turns out that the distribution of $p$
is uniform, between 0 and 1.
So, there is no reason to prefer $p=0.9$
in front of $p=0.2$.
Only small enough values of $p$ should lead
to the rejection of the null hypothesis.

The results for the same data using
$m=0.001$ (which increases $N$), also shown in Table \ref{t1},
are again in concordance with the scaling
hypothesis, being the smallest $p-$value
for this case larger than 0.14.
Even for $m=10^{-4}$ all the $p-$values are
above 0.2, except for some of the pairs
involving the set with $M\ge 2$.
The behavior of $d_{kl}$ when $m$ is changed,
which, following Ref. \refcite{Clauset} should
be a guide to chose the value of $m$ 
(although in a different type of test,
see next section),
is not clear in this case.

\section{Testing the Scaling Function}

A different statistical test regards the 
goodness of the fit applied to some data. 
For instance, we can ask whether
Eq. (\ref{gamma}) is a good approximation to the
empirical scaled distributions of inter-event times. 
Here, we will adapt the method of Ref. \refcite{Clauset}
to the kind of distributions that we are interested in.

First, a fit has to be performed.
A usual way of proceed in the case of long-tailed distributions
is to minimize the squared differences 
between the empirical density and the theoretical 
density in logarithmic scale;
however, this method 
shows some problems
and
involves the 
arbitrary estimation of the density;
other problems arises if one fits the cumulative distribution\cite{Clauset}.
In contrast, maximum likelihood estimation avoids these difficulties
by working directly with the ``raw'' data.

In order to be more general, let us consider
the distribution given by the probability density,
\begin{equation}
D(x) = \frac {\delta} {a\Gamma(\gamma/\delta,(m/a)^\delta)} 
\left(\frac a x \right)^{1-\gamma}
e^{-(x/a)^\delta},
\, \mbox{ for } x \ge m,
\label{gammagen}
\end{equation}
which constitutes the so called generalized gamma distribution,
with shape parameters $\gamma$ and $\delta$ and 
scale parameter $a$.
We consider $\gamma$ and $\delta$ greater than zero,
the opposite case can be considered as well
but then the function $\Gamma$ has to be replaced
by its complementary function (and multiplied by -1, as $\delta < 0$).
The cutoff value $m$ could be fixed to zero, 
but, as we have mentioned, for our data 
it is convenient to consider $m>0$.

The $n-$th moment of the distribution is given by
\begin{equation}
\langle x^n \rangle = a^n \,
\frac{\Gamma \left(\frac{\gamma+n}\delta,\frac{m^\delta}{a^\delta}\right)}
{\Gamma \left(\frac{\gamma}\delta,\frac{m^\delta}{a^\delta}\right)},
\end{equation}
for $\gamma >0$ and $\delta > 0$.
Notice 
that 
a particular case is given by the 
scaling function $f(x)$
appearing in Eq. (\ref{scaling}),
for which 
$\langle x \rangle \equiv 1$,
and only two of the three parameters are free;
nevertheless we will not make use of that restriction
for estimating the parameters.

The method of maximum likelihood estimation
is based on the calculation of the likelihood function $L$,
see Ref.  \refcite{Clauset}.
This is given by (or, in order to avoid dimensional problems, 
proportional to) the probability per unit of $x^N$ that 
the data set comes from a particular distribution,
given the values of its parameters
i.e.,
\begin{equation}
L(\gamma,\delta,a) = \frac{Prob[x_1, x_2, \dots x_N | \gamma,\delta,a]}
{d x_1, d x_2, \dots d x_N} \simeq \prod_{i=1}^N D(x_i | \gamma,\delta,a),
\end{equation}
where 
$N$ is the number of data 
and we make explicit the dependence of the 
probability density on its parameters.
The last step assumes that each value $x_i$ is independent 
on the rest. Naturally, this is not always the case
(we know that earthquake inter-event times are correlated%
\cite{Livina2,Corral_tectono,Lennartz})
and then the maximum likelihood method provides an estimation
of the distribution that generates the dataset in consideration
but it may be that the dataset is not representative 
of the process we are studying (due to correlations, the phase
space may not be evenly sampled).

It is more practical to work with the log-likelihood, $\ell$,
which is the logarithm of the likelihood;
dividing also by $N$,
\begin{equation}
\ell (\gamma,\delta,a) \equiv \frac {\ln L(\gamma,\delta,a)} N = 
\frac 1 N \sum_{i=1}^N \ln D(x_i | \gamma,\delta,a),
\end{equation}
which, notice, can be understood as a kind of estimator of
the entropy of the distribution from the available data
(with a missing $-1$ sign).
In the case of the generalized gamma distribution 
(\ref{gammagen}) it is
easy to get that
\begin{equation}
\ell (\gamma,\delta,a) = 
\ln \delta  -\ln \Gamma\left 
(\frac \gamma\delta,\left(\frac m a \right)^\delta\right)
+\gamma \ln \frac G a 
- \left(\frac {A(\delta)} a\right)^\delta,  
\label{logl}
\end{equation}
where we have omitted a term $-\ln G$ 
that is independent on the parameters of the distribution,
and we have introduced
$G$ as the geometric mean of the data, 
$\ln G \equiv (\sum \ln x_i)/N$, and
$A(\delta)$ as what we may call the $\delta-$power mean, 
$A(\delta) \equiv \sqrt[\delta] {\sum x_i^\delta/N}$ 
(which, in contrast to $G$, depends on the value of the parameter $\delta$;
for instance, for $\delta=1$, $A$ is the arithmetic mean, 
but for $\delta=-1$, $A$ is the harmonic mean).

The best estimate of the parameters would be that 
that maximizes the likelihood, or, equivalently, 
the log-likelihood.
The previous expression is too complicated to be maximized analytically,
and it is too complicated to differentiate even
(in fact, we would need to compute the derivative of the incomplete gamma 
function).
So, we will perform a direct numerical maximization
(in particular we will use the numerical routine 
{\tt amoeba} from Ref. \refcite{Press};
the function $\Gamma$ can be computed from the same
source using routines {\tt gammq} and {\tt gammln}).

Fixing $\delta\equiv 1$, from which we recover Eq. (\ref{gamma})
as a model of the distribution
(which yields only one free parameter and
has the advantage of being compatible with a Poisson process
in the limit of long times),
the resulting values of the parameters $\gamma$ and $a$,
obtained from maximum likelihood estimation,
are given in Table \ref{t2}.
In all cases, except for $M \ge 4$, 
and if the cutoff $m$ is not too small,
the values of $\gamma$ are close to 0.7.

Once we have obtained the estimators
of the parameters, we can ask about their meaning.
Maximum-likelihood estimation does not mean that
it is likely that the data comes from the 
proposed theoretical distribution, with those
parameters.
In fact, maximum likelihood can be minimum unlikelihood, 
i.e., we are taking the less bad option
among those provided by the {\it a priori} assumed
probability model.
In order to address this issue it is necessary to perform
a goodness-of-fit test.

Following Ref. \refcite{Clauset} we can 
employ again the Kolmogorov-Smirnov test, 
this time for one sample.
The KS statistic is, similarly as before
\begin{equation}
     d\equiv \mbox{max}_{\forall x} |P(x)-F(x)|
\end{equation}  
where $P(x)$ is the empirical cumulative distribution 
of the data,
defined in the previous section, and $ F(x)$
is the theoretical proposal.
For the distribution of Eq. (\ref{gammagen}),
\begin{equation}
F(x)\equiv \int_m^x D(x) dx = 1- \frac 
{\Gamma(\gamma/\delta ,(x/a)^\delta)} {\Gamma(\gamma/\delta ,(m/a)^\delta)}
\, \mbox{ for } x \ge m.
\label{cumul}
\end{equation}
The resulting values of $d$ for our problem are also shown in Table \ref{t2}.
Now we can apply the recipe of Clauset {\it et al.} in order to select the most
appropriate value of the cutoff $m$,
which consists in selecting the value which minimizes $d$.
Comparing between 0.003, 0.01, and 0.03,
it seems clear that we should chose $m=0.01$.

At this point
we could proceed as in the previous section,
using the formulas for the distribution of $d$.
However, that only would be right if
we were not estimating $F(x)$ from the data
(if we were comparing with a theory free of parameters for instance).
In order to know the distribution of the statistic $d$
when the data are generated by the model with 
the parameters obtained by maximum likelihood estimation, 
we will use Monte Carlo simulations.
Indeed, generating data from the theoretical distribution, 
we can repeat the whole process 
to obtain the statistical behavior of $d$
when the null hypothesis is true
(when the data come from the proposed theoretical distribution),
and we can do it many times, in order to get significant statistics.

Schematically, the process for the calculation of $p$
consists of the multiple iteration of the following steps:

\begin{enumerate}

\item Simulate synthetic data $s$ from the distribution given 
by Eq. (\ref{gamma}) using the parameters $\gamma$ and $a$
obtained before for the empirical data

\item Estimate the parameters $\gamma_s$
and $a_s$ by fitting the synthetic data $s$
to Eq. (\ref{gamma})
(proceeding in the same way as described above for the empirical data, 
see Eq. (\ref{logl}) and so on).

\item Evaluate the KS statistic for the distribution of synthetic
data $s$ [generated in (1) with parameters $\gamma$ and $a$]
and the theoretical distribution with parameters $\gamma_s$ and
$a_s$ [calculated in (2)], i.e., 
$d_s = \mbox{max}_{\forall x}|P_s(x|\gamma,a)-F(x|\gamma_s,a_s)|$

\end{enumerate}

We will obtain synthetic inter-event times from the gamma distribution
by generating a table of the cumulative distribution. 
As the probability of an event has to be the same independently on the
random variable we assign to that event, then, $u=F(x)$,
where $u$ is a uniform random number between 
zero and one 
and is also its own cumulative distribution.
We can calculate numerically the function $F(x)$
(thanks to the numerical recipes {\tt gammq} and {\tt gammln}\cite{Press}),
but are unable to calculate its inverse
(at least, in a reasonable computer time), 
so we will tabulate the values of $F(x)$, 
for selected values of $x$
in log scale
(this is to deal with the multiple time scales that 
appear in the process, described by Eq. (\ref{gamma})
or (\ref{cumul}) when $\gamma < 1$ and $m \ll 1$).
To be concrete, 
$
x(k)=m e^{\alpha k}
$,
where $k=0,1,\dots$ and $\alpha$ is just a constant.
Then, when a uniform value $u$ is generated 
we can obtain the corresponding value of $x$
by looking at the table and interpolating (or extrapolating)
using the closest values of $u(k)=F(x(k))$.

\begin{figure}[bt]
\centerline{
(a)
\psfig{file=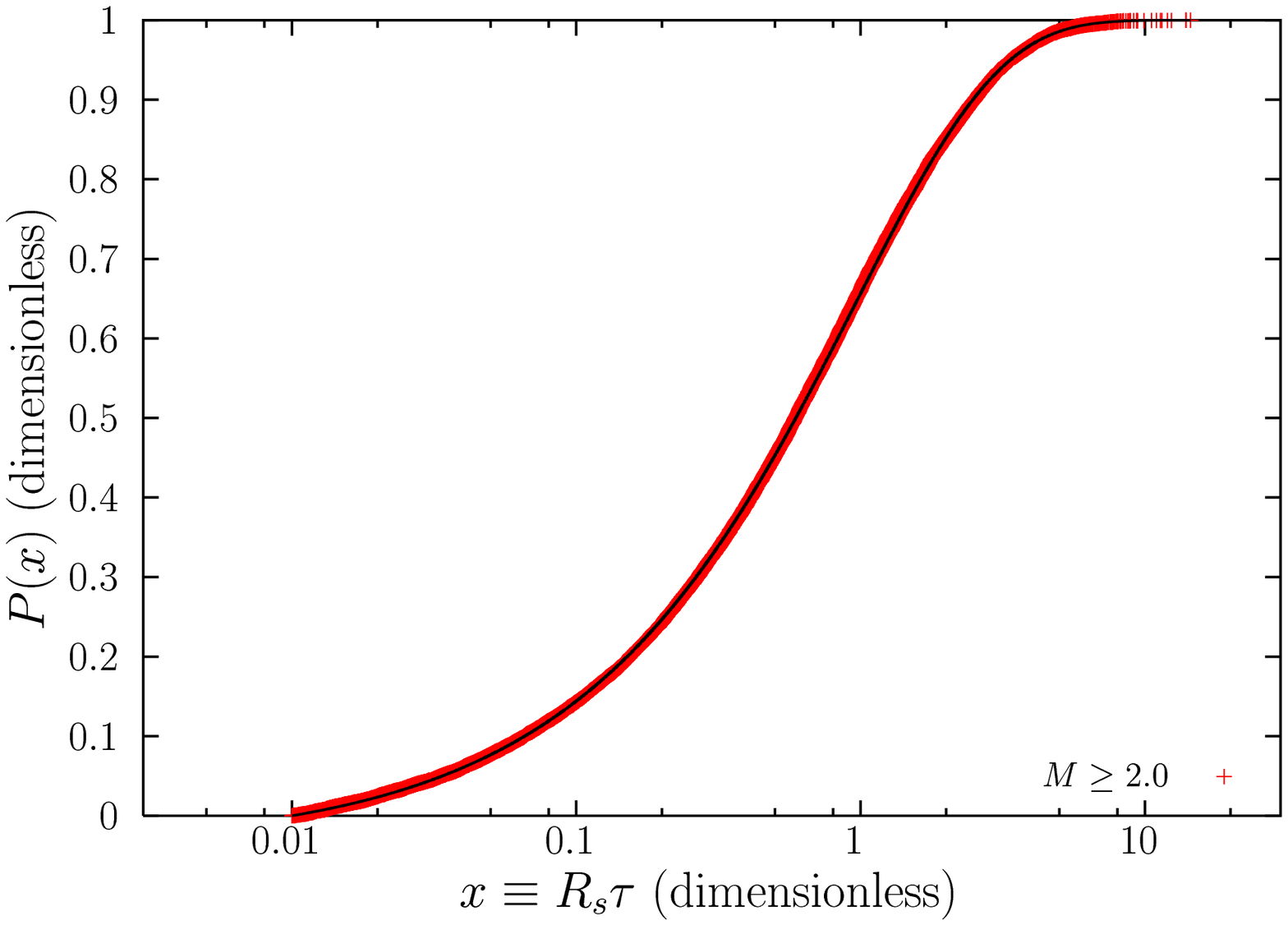,height=6cm}
(b)
\psfig{file=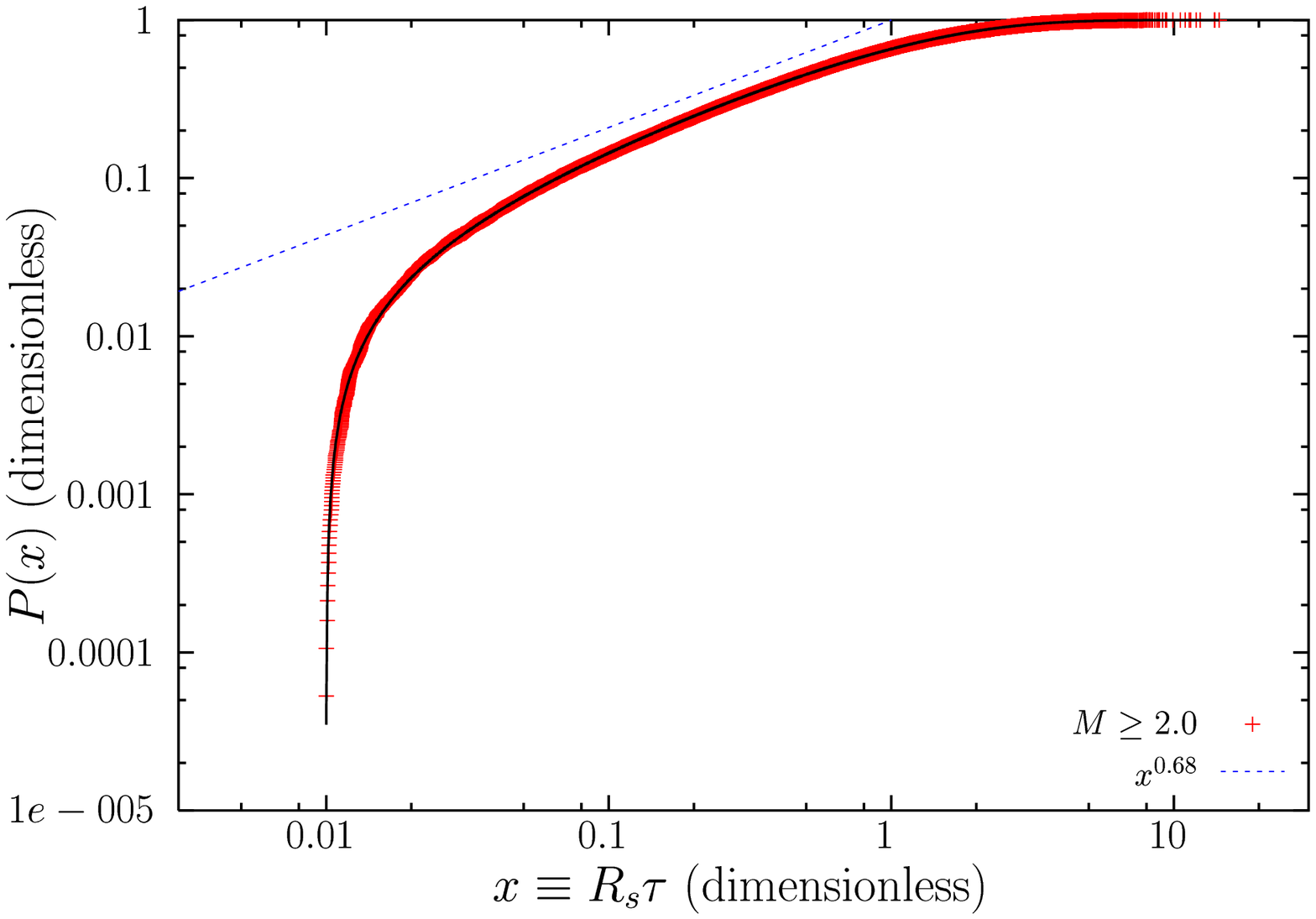,height=6cm}
}
\centerline{
(c)
\psfig{file=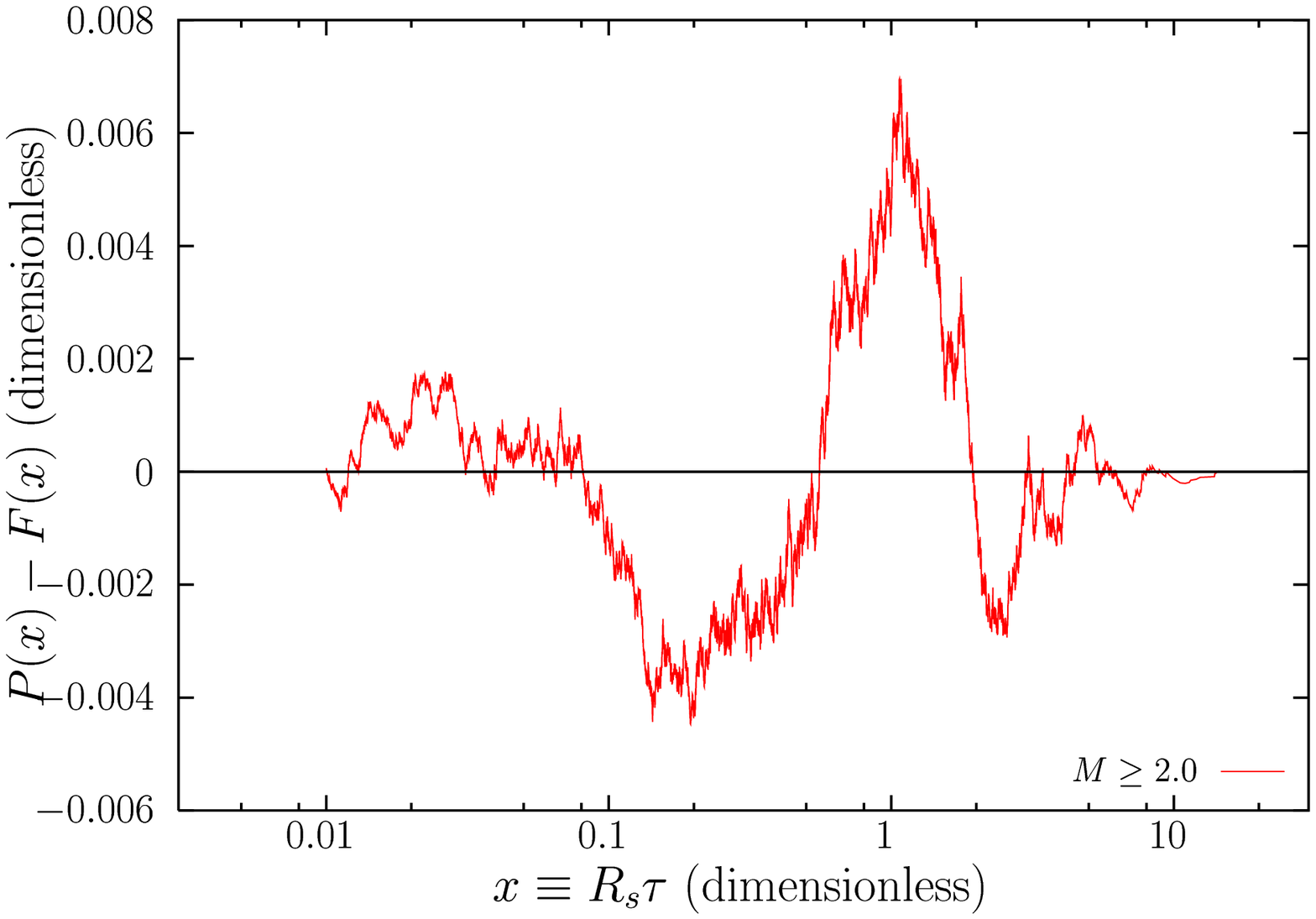,height=6cm}
(d)
\psfig{file=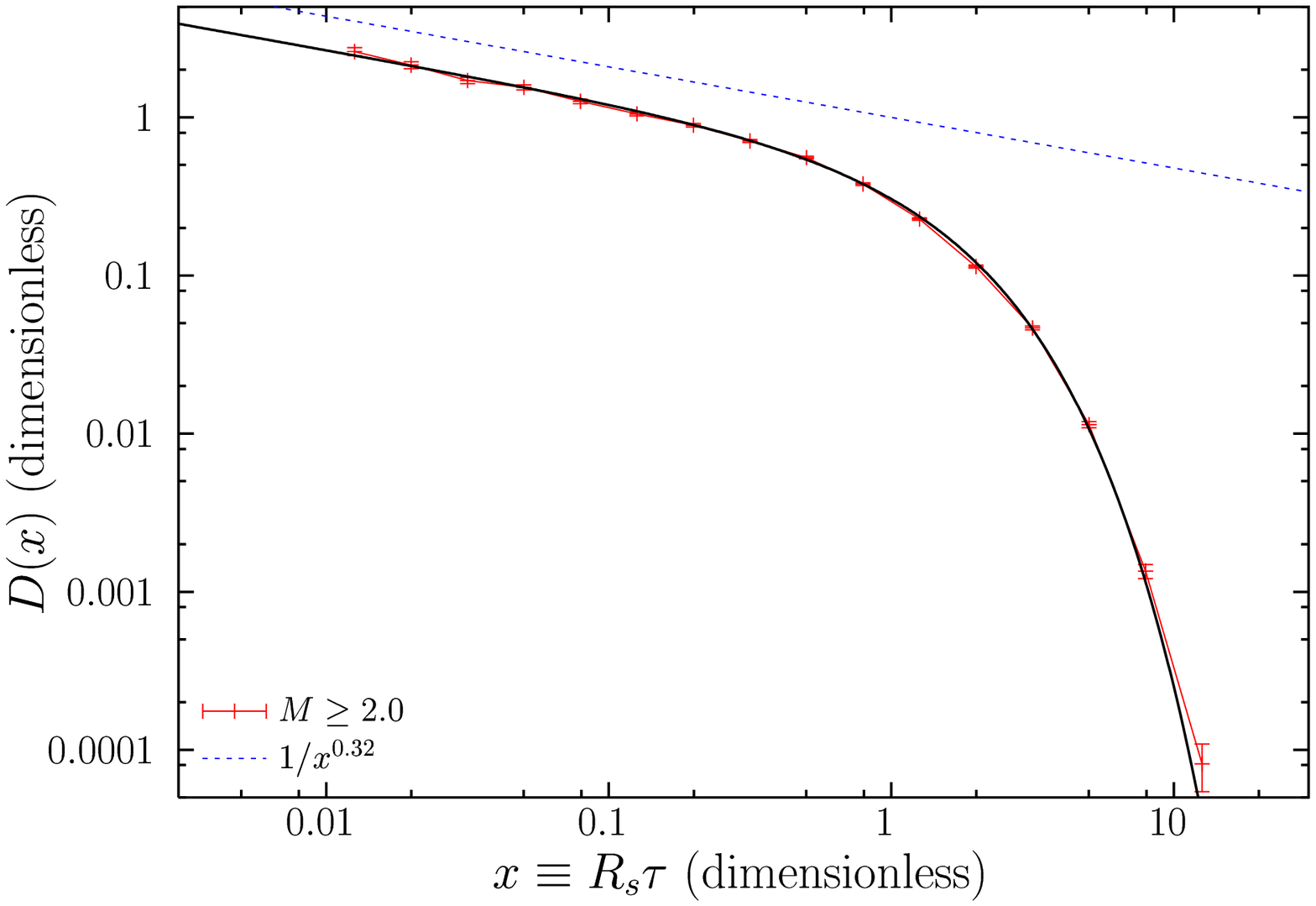,height=6cm}
}
\vspace*{8pt}
\caption{
(a)
Rescaled inter-event-time cumulative distribution
for Southern California stationary seismicity with $M\ge 2$,
fixing minimum $x-$value $m=0.01$, together with the 
fit obtained by maximum likelihood estimation.
The $p-$value corresponding to the KS statistic,
determined by Monte Carlo simulation turns out to be
as small as 0.032, although the fit is visually
acceptable.
(b)
Same as before, in log-log scale,
together with a pure power law with the same exponent.
(c)
Difference between the distribution and its fit,
$P(x)-F(x)$, which yields the KS statistic
when its absolute value is maximized.
(d) 
The corresponding probability density, estimated with
5 bins per decade, for comparison.
Also the best fit and a pure power law are shown.
\label{onesample}
}
\end{figure}

For the case of our interest,
the $p-$values calculated in this way, 
using 1000 randomly generated samples
(which yield an uncertainty of about 3 \% in $p$),
are included in Table \ref{t2}.
Taking $m=0.01$
(the value arising by the application of Clauset {\it et al.}
recommendation), 
we cannot reject the hypothesis
that the data set comes from the theoretical distribution
with maximum likelihood parameters, except for 
$M \ge 2$, which yields $p=0.032$, 
which is beyond the usual onset of acceptance of the null
hypothesis, $p=0.05$.
Figure \ref{onesample} illustrates the reason of the rejection.
Indeed, although the theoretical distribution
is very close to the empirical one, 
the difference is large enough for the high number of data involved.
Although Eq. (\ref{Q}) is not valid in this case,
we can use it as an approximation and see 
how for large $N$ the statistic $d$ scales
as $1/\sqrt{N}$ ($N_e=N$ here).
As the mode of the distribution $Q$ in Eq. (\ref{Q}) 
is around 0.735 and practically all the probability
is contained below 2, this means that
we can expect $d < 2/\sqrt{N}$.
So, for large $N$, $d$ tends to zero, 
and the KS test is able to detect any small difference
between the proposed theoretical distribution and the 
``true'' distribution.
This means that the test is not adequate if we are just
interested to find an approximation to the true distribution, 
as only the ``true'' distribution is not rejected 
for a sufficient number of data.
For comparison, we show in Table \ref{t3} 
the results for an exponential scaling function,
which is clearly rejected except for $M\ge 4$.

\section{Discussion}

As another alternative, note that we have tested separately
the validity of the scaling law and
the adequacy of the scaling function given by Eq. (\ref{gamma}).
We could take advantage of the scaling behavior to 
fit and test the goodness of fit of the scaling function.
For instance, we could combine all rescaled data sets
(for all values of the minimum magnitude)
and proceed as in Sect. 3 for this combined data set.
The problem is that, by virtue of the Gutenberg-Richter law,
when the minimum magnitude is raised in one unit, the number of
events decreases by a factor 10, and therefore, data sets with large
minimum magnitudes are under-represented.
Perhaps we could just truncate the samples in order that all of them had 
the same number of data, but that would lead to a tremendous
wasting of information.

The surprising character of the scaling law (\ref{scaling})
when the scaling function is not exponential has lead to some criticisms 
by Molchan\cite{Molchan} and Saichev and Sornette\cite{Saichev_Sornette_times}.
The latter authors propose that, for the so called ETAS model,
the scaling law is not valid, and one has a very slow variation
of the inter-event-time distribution when the magnitude threshold is raised.
Although we have tested that the scaling law is consistent with the data
within statistical significance, this does not mean that we
should reject Saichev and Sornette result.
Nevertheless, the simplicity of the scaling hypothesis
makes it the most adequate model for seismicity, at least as
a null model to contrast with other hypothesis.
On the other hand, other seismicity models have been recently proposed,
which, in contrast to the ETAS model, are fully scale invariant and one would
expect that are characterized by scaling inter-event-time distributions%
\cite{Vere_Jones,Saichev_Sornette_Vere_Jones,Lippiello,bass1,bass2}.

Saichev and Sornette also provide a pseudo-scaling function
to which inter-event-time distributions can be fit\cite{Molchan,Saichev_Sornette_times}.
In principle, the very same procedure used in our paper can be
applied directly in order to fit the parameters of the Saichev-Sornette
function and test the goodness-of-fit of the outcome.
It is expected that the use of this new function, 
which has more parameters than Eq. (\ref{gamma})
and models better the left tail of the distribution,
could lead to the reduction of the cutoff $m$ 
above which the functions are fit.
We leave this task for future research.

\begin{table}[ph]   
\tbl{KS statistic $d$ (below the diagonal) and corresponding $p-$value
(above diagonal, in percentage) 
for rescaled Southern-California stationary-seismicity
inter-event-time distributions,
with lower cutoffs $m=0.01$ (top) and $m=0.001$ (bottom).
The scaling hypothesis cannot be rejected.
\label{t1}
}
{\begin{tabular}{@{}lr|rrrrr@{}} \Hline 
\\[-1.8ex] 
$m=0.01\phantom{0}$
& $N$ & $M \ge 2.0$ &  $M \ge 2.5$ &  $M \ge 3.0$ &  $M \ge 3.5$ &  $M \ge 4.0$ \\ \hline
$M \ge 2.0$ & 18870 &            -        &   26.1\%     &    36.0\%    &     97.3\%  &      21.3\% \\
$M \ge 2.5$ & 4953 &           0.016    &       -       &   63.2\%    &    86.3\%   &     18.3\% \\
$M \ge 3.0$ & 1184 &           0.028    &      0.024   &        -       &   95.7\%    &    22.8\% \\
$M \ge 3.5$ & 309 &           0.028    &      0.035   &       0.032   &        -       &   20.8\% \\
$M \ge 4.0$ & 70 &          0.125     &    0.129     &    0.126      &   0.138       &    -  \\
\Hline \\[-1.8ex] 
\end{tabular}}

{\begin{tabular}{@{}lr|rrrrr@{}} \Hline 
\\[-1.8ex] 
$m=0.001$
& $N$ & $M \ge 2.0$ &  $M \ge 2.5$ &  $M \ge 3.0$ &  $M \ge 3.5$ &  $M \ge 4.0$ \\  \hline
$M \ge 2.0$ & 19821 &                     -       &    41.6\%      &  43.8\%       & 14.3\%      & 31.9\%\\
$M \ge 2.5$ & 5187 &                     0.014  &         -     &     57.1\%    &    29.6\%    &    31.1\%\\
$M \ge 3.0$ & 1268 &                     0.025  &        0.024 &          -    &      68.0\%  &      33.5\%\\
$M \ge 3.5$ & 340 &                     0.062  &        0.054  &        0.044&           -  &        28.1\%\\
$M \ge 4.0$ & 76 &                     0.108  &        0.110  &        0.110&         0.124&           -  \\
\Hline \\[-1.8ex] 
\end{tabular}}
\end{table}

%

\begin{table}[ph]   
\tbl{Maximum likelihood parameters $\gamma$ and $a$,
KS statistic $d$ and corresponding $p-$value
(in percentage, determined by Monte Carlo simulation) 
for rescaled Southern-California stationary-seismicity
inter-event-time distributions,
using several values of the minimum value $m$.
\label{t2}
}
{\begin{tabular}{@{}l|rrrrr@{}} \Hline 
\\[-1.8ex] 
$m=0.03\phantom{0}$
     & $N$&$\gamma$ & $a$  & $d$   & $p-$value \\ \hline
$M \ge 2.0$ & 18009 & 0.68 &1.35 & 0.008 &   1.2\% \\
$M \ge 2.5$ &  4669 & 0.67 &1.38 & 0.007 &  84.0\% \\
$M \ge 3.0$ &  1122 & 0.73 &1.29 & 0.021 &  25.7\% \\
$M \ge 3.5$ &   287 & 0.79 &1.22 & 0.034 &  55.7\% \\
$M \ge 4.0$ &    69 & 0.89 &1.07 & 0.089 &  16.3\% \\
\Hline \\[-1.8ex] 
\end{tabular}}

{\begin{tabular}{@{}l|rrrrr@{}} \Hline 
\\[-1.8ex] 
$m=0.01\phantom{0}$   
            & $N$&$\gamma$ & $a$  & $d$   & $p-$value \\ \hline
$M \ge 2.0$ & 18870 & 0.68 & 1.41 & 0.007 &  3.2\%  \\
$M \ge 2.5$ & 4953 &  0.64 & 1.50 & 0.009 & 37.3\%  \\
$M \ge 3.0$ & 1184 &  0.69 & 1.41 & 0.021 & 21.8\%  \\
$M \ge 3.5$ & 309 &   0.67 & 1.45 & 0.029 & 73.6\%  \\
$M \ge 4.0$ & 70 &    0.95 & 1.05 & 0.082 & 29.3\%  \\
\Hline \\[-1.8ex] 
\end{tabular}}

{\begin{tabular}{@{}l|rrrrr@{}} \Hline 
\\[-1.8ex] 
$m=0.003$   & $N$&$\gamma$ & $a$  & $d$   & $p-$value \\ \hline
$M \ge 2.0$ & 19466  & 0.65 & 1.51 &  0.009 &    0.0\% \\
$M \ge 2.5$ &  5102 &  0.62 & 1.57 &  0.011 &   11.2\% \\
$M \ge 3.0$ &  1247 &  0.59 & 1.65 &  0.029 &    1.4\% \\
$M \ge 3.5$ &   328 &  0.56 & 1.74 &  0.046 &    9.0\% \\
$M \ge 4.0$ &    74 &  0.72 & 1.37 &  0.103 &    5.8\% \\
\Hline \\[-1.8ex] 
\end{tabular}}

\end{table}


\begin{table}[ph]   
\tbl{Same as the previous table,
for the exponential distribution ($\gamma  \equiv 1$).
\label{t3}
}
{\begin{tabular}{@{}l|rrrrr@{}} \Hline 
\\[-1.8ex] 
$m=0.01$
            & $N$&$\gamma$ & $a$  & $d$   & $p-$value \\ \hline
$M \ge 2.0$ & 18870& 1 & 0.99 & 0.072 &    0.0\% \\
$M \ge 2.5$ & 4953 & 1 & 0.99 & 0.084 &    0.0\% \\
$M \ge 3.0$ & 1184 & 1 & 1.00 & 0.077 &    0.0\% \\
$M \ge 3.5$ &  309 & 1 & 1.00 & 0.079 &    0.4\% \\
$M \ge 4.0$ &   70 & 1 & 1.00 & 0.077 &   59.5\% \\
\Hline \\[-1.8ex] 
\end{tabular}}
\end{table}


\section*{Acknowledgements}

The author has been benefited by discussions with A. Deluca and
R. D. Malmgren, and appreciate the generosity of Shearer
{\it et al.} to make the results of their research 
publicly available.
This research has been part of the Spanish projects
FIS2006-12296-C02-01
and
2005SGR-0087.

\section*{References}




\end{document}